\documentclass[conference]{IEEEtran}

\usepackage[english]{babel}

\usepackage{amsmath}
\usepackage{graphicx}

\usepackage{titlesec}
\titlespacing*{\section}{0pt}{0.8ex plus 0.3ex minus 0.2ex}{0.6ex plus 0.2ex}
\titlespacing*{\subsection}{0pt}{0.7ex plus 0.2ex minus 0.2ex}{0.5ex plus 0.2ex}

\title{Availability Attacks Without an Adversary:\\ Evidence from Enterprise LANs}

\author{
\IEEEauthorblockN{Rajendra Paudyal, Rajendra Upadhyay, Al Nahian Bin Emran, Lisa Donnan, Duminda Wijesekera}
\IEEEauthorblockA{
Mason Innovation Labs, George Mason University, Fairfax, VA 22030, USA\\
\{rpaudyal\textbar{} rupadhya\textbar{} abinemra\textbar{} ldonnan\textbar{} dwijesek\}@gmu.edu
}
}

\makeatletter
\def\@IEEEtitleblockvspace{0.1\baselineskip}
\makeatother

\begin{document}
\maketitle

\begin{abstract}
Denial-of-Service (DoS) conditions in enterprise networks are commonly attributed to malicious actors. However, availability can also be compromised by benign non-malicious insider behavior. This paper presents an empirical study of a production enterprise LAN that demonstrates how routine docking and undocking of user endpoints repeatedly trigger rapid recalculations of the control plane of the Rapid Spanning Tree Protocol (RSTP)~\cite{cisco_rstp_change}. Although protocol-compliant and non-malicious, these events introduce transient forwarding disruptions of approximately 2-4 seconds duration that degrade real-time streaming (voice and video) services while remaining largely undetected by conventional security monitoring. We map this phenomenon to the NIST and MITRE insider threat frameworks, characterizing it as an unintentional insider-driven availability breach, and demonstrate that explicit edge-port configuration effectively mitigates the condition without compromising loop prevention.
\end{abstract}
\section{Introduction}
\label{sec:introduction}
The security of enterprise information systems is framed around the CIA triad of confidentiality (C), integrity (I), and availability (A). Although confidentiality and integrity violations often receive the most attention due to their association with data breaches and regulatory compliance, availability is equally critical, particularly in modern enterprises that rely heavily on real-time digital communication~\cite{nwOutage}. A brief interruption in service quality and availability can stop business operations, degrade productivity, and undermine trust in the IT infrastructure.
Modern enterprises rely on real-time applications, such as voice over IP (VoIP), video conferencing, collaborative whiteboards, and cloud-hosted productivity platforms. Unlike best-effort data services, these applications are highly sensitive to packet loss, delay, jitter, and even short-duration interruptions. A network disruption lasting only a few seconds can freeze video streams, trigger session renegotiation, and cause noticeable degradation of the user experience. From a security perspective, such events represent violations of availability guarantees and therefore fall within the scope of denial-of-service conditions.

Today's office environments within enterprises have undergone a significant transformation. The widespread adoption of hybrid work models and flexible seating arrangements has led to the proliferation of user-operated peripherals, such as USB-C docking stations and monitors with integrated Ethernet interfaces. These devices are designed for convenience and are connected and disconnected throughout the workday. They are generally perceived as benign accessories with little or no security impact. To serve such application connectivity, today's enterprise networks continue to rely on Layer 2 control-plane protocols, such as Rapid Spanning Tree Protocol (RSTP), which were designed under the assumption of relatively static topologies and infrequent link changes~\cite{cisco_rstp_change}. RSTP is enabled by default on enterprise switches to prevent switching loops and broadcast storms. Due to RSTP's rapid convergence properties, typically on the order of a few seconds, they are considered sufficient for most applications and are assumed to be transparent to users. This paper challenges this assumption. We present a detailed empirical study of a production enterprise's LAN, in which the normal use of docking stations by employees unintentionally triggered repeated RSTP recalculations. These recalculations caused transient forwarding interruptions lasting 2 to 4 seconds, resulting in random but recurring interruptions to real-time voice and video services. Importantly, no malicious activity, misconfiguration, or policy violation was involved. The disruptions arose solely from the interaction between the default behavior of the Layer-2 protocol and the routine actions of the user.
This paper examines one such risk and demonstrates how benign insider activity can unintentionally induce protocol-compliant, yet disruptive, control-plane behavior.

\subsection{Motivation}
\label{ssec:motivation}
This work is motivated by a growing gap between how Layer-2 control protocols are designed and how modern enterprise endpoints behave. Protocols such as RSTP assume that network topology changes are rare and mostly caused by failures or administrative actions. In today’s enterprise networks, however, routine user behavior such as frequently docking and undocking laptops causes repeated and unintentional link changes at the access layer. Current enterprise security and monitoring systems focus mainly on malicious attacks and therefore treat brief availability disruptions as minor operational issues rather than security events. This is problematic for real-time services, where even short control-plane interruptions can have a noticeable impact. As a result, availability risks caused by benign user behavior are difficult to detect and mitigate. This work aims to study how everyday endpoint actions interact with Layer-2 protocols to create overlooked availability risks in enterprise networks.

\subsection{Contribution}
This paper makes the following contributions.
\begin{itemize}
    \item We present the first empirical study demonstrating how benign insider behavior can repeatedly trigger Layer-2 control-plane recalculations in production enterprise networks.
    \item We show that routine docking and undocking events can induce protocol-compliant RSTP behavior that results in measurable availability degradation for real-time applications.
    \item We formally map this phenomenon to NIST and MITRE insider threat frameworks, framing it as an unintentional, insider-driven denial-of-service condition.
    \item We identified a previously unrecognized mismatch between the assumptions of the Layer-2 protocol and the modern behavior of the enterprise endpoint.
    \item We suggest explicitly configuring edge ports using equivalent PortFast settings to ensure that such ports are treated as host only interfaces and therefore excluded from spanning tree topology calculations.
\end{itemize}

\section{Background and Related Work}
\label{sec:background+related Work}

\begin{figure}[!ht]
\label{topo}
\centering
\includegraphics[width=0.95\linewidth]{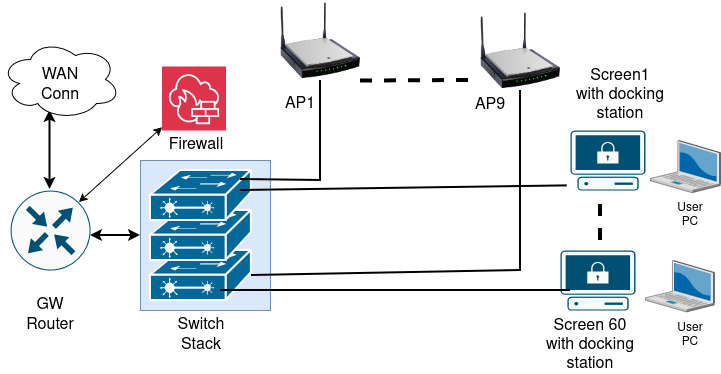}
\caption{\label{fig:topo}Enterprise access-layer topology. }
\end{figure}

This section reviews related work in four areas: insider threats and accidental insiders, availability and denial-of-service issues, Layer-2 control-plane stability, and the impact of network disruptions on real-time enterprise applications. We conclude by highlighting a clear research gap that motivates this study.

\subsection{Insider Threats and Accidental Insiders}
\label{ssec:insider}

Insider threats have long been recognized as a significant risk to organizational security due to the legitimate access of insiders to systems and data. The early foundational work by Shaw and Fischer~\cite{ref3} and later by Cappelli et al.~\cite{ref1} established comprehensive insider threat taxonomies, emphasizing that insiders can cause harm maliciously or unintentionally. The  U.S. Computer Emergency Preparedness Team (CERT) works within the Department of Homeland Security (DHS) and further formalized the concept of \emph{accidental insider}, identifying unintentional actions such as mistakes, lack of awareness, or misuse of tools as major contributors to security incidents~\cite{cisco_rstp_change}.
Subsequent research expanded insider threat analysis beyond purely malicious behavior. Greitzer et al.~\cite{ref2} introduced predictive modeling approaches that consider behavioral and contextual factors rather than intent alone. These studies highlight that intent is often difficult to infer and that security impact should be a primary consideration when assessing insider-related risk.

Despite this recognition, most insider threat research continues to focus on confidentiality and integrity violations, such as data exfiltration, fraud, or sabotage. Availability centric insider threats, particularly those arising from benign user behavior, remain underrepresented in both academic literature and enterprise security tooling. This imbalance creates blind spots in security architectures, where disruptions caused by legitimate insiders may not be treated as security incidents.

\subsection{Availability and Denial-of-Service as Security Incidents}
\label{ssec:DoS}
Availability is explicitly defined as a core security objective in the CIA triad and is formally addressed in multiple NIST publications. NIST SP 800-53~\cite{ref4} includes availability focused controls such as denial-of-service protection, while NIST SP 800-61~\cite{ref5} classifies denial-of-service incidents based on impact rather than attacker intent. According to this definition, any event that significantly degrades service availability qualifies as a security incident, regardless of whether it originates from an external attacker, a misconfiguration, or internal user behavior.
Research on denial-of-service attacks has traditionally emphasized external adversaries and large-scale distributed attacks. Mirkovic and Reiher~\cite{ref6} presented a widely cited taxonomy of DoS attacks and defense mechanisms, focusing on attack vectors, traffic characteristics, and mitigation strategies. Although this work established the foundational understanding of DoS, it implicitly assumes an adversarial attacker model and does not consider insider-driven or protocol-induced availability degradation.
Recent work highlights the growing importance of availability in enterprise and national security networks. Availability degradation has been observed in cellular systems, where 5G backhaul delays can impair time-critical responses~\cite{Upadhyay2025UAV}. SDN-based studies show that traffic and load-management decisions, including backbone congestion reduction~\cite{rp} and fuzzy-logic-based real-time load balancing~\cite{dc}, directly affect network availability. Clarke~\cite{ref7} noted that availability failures can be as damaging as data breaches. However, accidental and protocol-driven availability failures remain largely unexplored.

\subsection{Layer-2 Control-Plane Protocols and Stability}
\label{ssec:L2}
Layer-2 loop prevention mechanisms, particularly Spanning Tree Protocol (STP) and its successor RSTP, are foundational components of Ethernet-based enterprise networks. IEEE 802.1w~\cite{ref9} defines RSTP as an improvement over legacy STP, providing faster convergence, typically within a few seconds after topology changes. Classic networking literature, such as Perlman’s seminal work on interconnections~\cite{ref10}, treats spanning tree protocols as correctness mechanisms designed to ensure loop free topologies. These works assume relatively static network topologies with infrequent changes caused primarily by failures or administrative actions. Operational studies and vendor documentation, such as Cisco’s RSTP white papers~\cite{ref11}, focus on best practices in configuration and convergence behavior, but do not examine the application layer or the security implications of frequent topology changes. Huc et al.~\cite{ref12} analyze stability issues in Ethernet spanning tree protocols, demonstrating that repeated topology changes can cause instability; however, their work focuses on protocol dynamics rather than real world enterprise usage patterns or insider-driven triggers. Notably absent from previous work is an examination of human-triggered access layer topology changes caused by modern peripherals, such as docking stations. Existing studies do not consider how routine user behavior might repeatedly invoke control plane recalculations in production environments.

\subsection{Sensitivity of Real Time Applications to Network Disruptions}
\label{ssec:sensitivity}

The sensitivity of time critical voice and video applications to network impairments is well documented. ITU-T Recommendation G.114~~\cite{ref13} defines strict latency and jitter thresholds for acceptable voice quality, noting that even short interruptions can significantly degrade user experience. Schulzrinne et al.~\cite{ref14}, in the RTP specification, emphasize that real time transport protocols lack the retransmission and congestion recovery mechanisms of TCP, making them particularly vulnerable to packet loss and brief outages.
Enterprise application vendors have echoed these findings. Microsoft’s official documentation for Teams~\cite{ref15} specifies that even short periods of packet loss or increased latency can result in call freezes, lost audio, or poor video quality. These characteristics make real time applications effective indicators of underlying network instability, even when traditional performance metrics appear normal.
While these studies establish the sensitivity of real time applications, they generally assume that disruptions originate from congestion, wireless issues, or external attacks. The impact of Layer-2 control plane events, especially those triggered by legitimate insider actions, has received little attention.

Previous work establishes that insider threats include not only malicious actors, but also unintentional or accidental insiders whose legitimate actions may negatively impact organizational systems~\cite{ref1,ref2,ref4}. In addition, availability degradation is widely recognized as a security incident irrespective of adversarial intent, with denial-of-service conditions classified according to impact rather than motivation~\cite{ref5, ref6}. In the network control plane, RSTP has been standardized to provide fast convergence and loop prevention in Ethernet networks; however, its design assumes relatively infrequent topology changes, typically associated with failures or planned maintenance events~\cite{ref9, ref10}. Existing studies also show that real time enterprise applications, including voice and video conferencing systems, are highly sensitive to short network disruptions, where interruptions of only a few seconds can lead to perceptible service degradation~\cite{ref13,ref14,ref15}. Despite these observations, the literature lacks empirical analyses of how benign insider behavior interacting with default Layer-2 control protocols can cause persistent availability degradation in enterprise environments. Specifically, there is limited work framing user triggered RSTP recalculations as insider-driven denial-of-service conditions or formally mapping such behavior to established NIST and MITRE insider threat frameworks. This paper addresses these gaps through a real world case study that links routine docking station usage to repeated RSTP recalculations and measurable degradation of real-time services, revealing a previously under recognized class of unintentional insider-driven availability attacks.

\section{System Model and Network Architecture}
\label{sec:systemModel}

The studied environment includes multiple branches that are housed in different locations. Each site is equipped with Cisco Meraki MS250 access switches, with RSTP enabled globally by default on all interfaces, including access ports. Network connectivity to external networks is provided through a Meraki gateway router in each branch. All access ports operate in the same VLAN 1 and participate in a single RSTP instance managed at the switch stack level. Figure~\ref{fig:topo} shows three access switches that operate as a single logical stack, with user laptops connected via docking stations and access points connected directly to access ports. RSTP is enabled on all ports by default. 

User endpoints consist primarily of employee laptops connected either through a wireless AP or connected to a switch via a wired channel indirectly through docking stations. Connections happen mostly through the wireless channel, but when the user wants to use the screen, the user's laptop gets connected to the network through the wired channel. Many of these docking devices implement internal switching or bridging functionality, leading access switches to interpret them as intermediate Layer-2 devices rather than simple end hosts. Consequently, normal user actions, such as connecting or disconnecting a docking station during the workday, introduce physical link-up and link-down events at the access layer. These frequent docking and undocking events trigger RSTP port role transitions and topology change notifications. These control plane events form the basis for the availability degradation analyzed in this study.

\subsection{Threat Model and Insider Threat Mapping}
\label{ssec:mappinThreatModels}

The threat model assumes an authorized employee acting as an unintentional insider with legitimate, least privilege access to the enterprise network. Routine user actions, such as docking and undocking peripherals, trigger repeated RSTP recalculations in the Layer-2 control plane, resulting in a transient but recurring availability degradation. Although malicious intent is not involved, the resulting impact qualifies as a security incident under NIST SP 800-61, maps to the Denial of Service (TA0040) impact tactic in the miter ATT\&CK framework, and corresponds to the Denial of Service category in STRIDE, making it functionally equivalent to a low-rate internal Denial of Service condition~\cite{MITRE, NIST80060, STRIDE}. Table~\ref {tab:threat-model} shows the threat model and the insider threat mapping of the study.

\begin{table}[!ht]
\centering
\caption{Threat Model and Insider Threat Mapping}
\label{tab:threat-model}
\renewcommand{\arraystretch}{1.3}
\begin{tabular}{|p{2.2cm}|p{4.4cm}|}
\hline
\textbf{Threat Dimension} & \textbf{Description} \\
\hline
Threat Actor & Authorized employee with legitimate access to the enterprise network \\
\hline
Access Level & Least-privilege user access to access-layer network resources \\
\hline
Intent & Non-malicious, routine work-related behavior \\
\hline
Insider Classification & Accidental insider, as defined by NIST insider threat taxonomy \\
\hline
Threat Vector & Layer-2 network control plane \\
\hline
Trigger Mechanism & Routine docking and undocking of peripherals connected to access ports \\
\hline
Control-Plane Mechanism & Repeated Rapid Spanning Tree Protocol (RSTP) recalculations \\
\hline
Security Impact & Transient but recurring degradation of network availability \\
\hline
NIST Mapping & Availability impact classified as a security incident (NIST SP 800-61) \\
\hline
MITRE ATT\&CK Mapping & Impact tactic: Denial of Service (TA0040) \\
\hline
STRIDE Mapping & Denial of Service \\
\hline
Threat Characterization & Functionally equivalent to a low-rate internal denial-of-service condition \\
\hline
\end{tabular}
\end{table}

\section{Measurement methodology}
\label{sec:measurement}

The measurement methodology adopted in this study follows the principles of risk identification and impact analysis outlined in NIST SP 800-30, with a specific focus on the availability risk arising from internal operational behavior. Data collection was carried out in three branch offices in the enterprise with identical network configurations, ensuring the consistency and reproducibility of the observations. Measurements were conducted over multiple business days, capturing repeated docking and undocking events under normal operating conditions.

\begin{figure}[!ht]
\label{logwithrstp}
\centering
\includegraphics[width=1\linewidth]{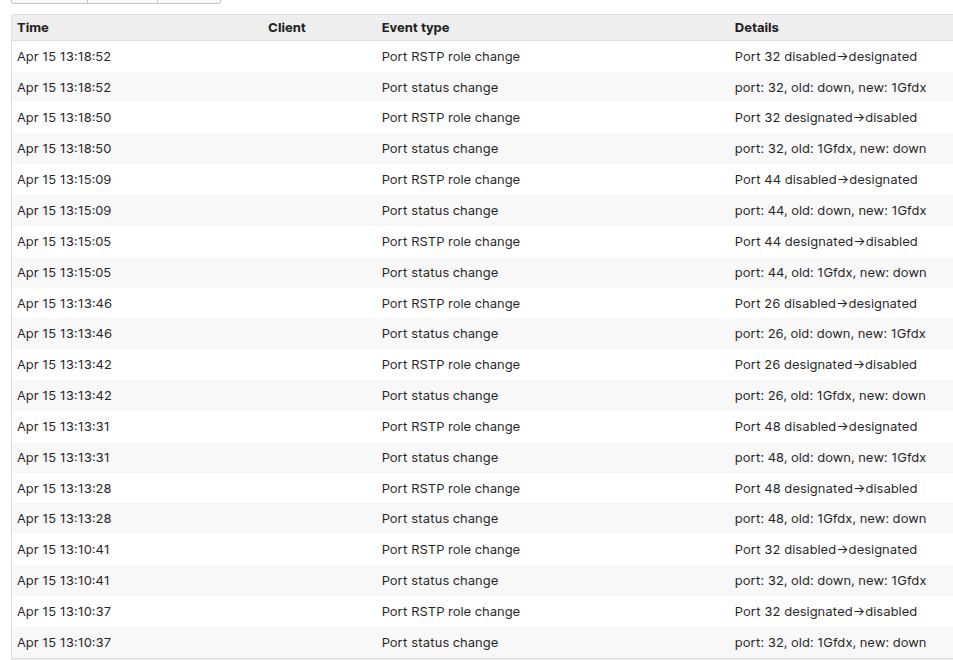}
\caption{\label{fig:rstp} RSTP/port-state changes log. For privacy reasons, we excluded information like IP, switch name, and location metadata from the snap. }
\end{figure}

First, control-plane and interface-level logs were collected from Cisco Meraki MS250 access switches, including RSTP events, port state transitions, and topology change notifications. These logs provided precise timestamps for link-up, link-down, and spanning tree convergence events. Second, event correlation was performed by aligning docking and undocking times captured through switch interface logs and user activity records with corresponding RSTP recalculations. This correlation enabled direct attribution of control-plane events to specific user actions.

Third, application-level behavior was observed by monitoring ongoing video conferencing and VoIP sessions during normal business operations. Quality degradation was identified through observable indicators such as audio clipping, video freezes, and session renegotiations reported by users, as well as packet loss and jitter metrics where available. Finally, controlled experiments were conducted outside peak hours by repeatedly docking and undocking selected endpoints while maintaining active real-time application sessions. These experiments were designed to isolate causality, eliminate confounding variables, and verify that the observed availability degradation consistently followed docking related RSTP events.

\subsection{Experimental Results}
\label{ssec:experimenatlResults}
Across all measured sites, each docking or undocking event consistently triggered RSTP port-state transitions on the affected access interface, typically progressing from Designated to Disabled and back to Designated, with convergence times of approximately 2 to 4 seconds. Figure~\ref{fig:rstp} illustrates access-switch logs showing the corresponding port-state transitions and topology change events. During these convergence intervals, real-time applications experienced visible degradation, including video freezes, audio dropouts, and increased packet loss, while TCP-based applications recovered transparently through retransmission and buffering, masking the underlying instability. In particular, the impact was not limited to the flapping port: multiple hosts and access points connected to the same switch exhibited simultaneous service disruption, indicating that control-plane recalculations and associated data-plane effects influenced forwarding behavior beyond a single interface. The repeatability of these observations across sites and controlled experiments confirms a causal relationship between routine docking activity, protocol recalculations, and transient availability degradation in real-time enterprise services.

\section{Root Cause Analysis}
\label{sec:rootCause}
The availability disruption observed is rooted in the transition of the Spanning Tree state machine. When a docking station induces a link-state flap, it triggers a Topology Change Notification (TCN). In a standard RSTP environment, the total convergence time $T_{conv}$ during which the data plane is stalled can be modeled as:

\begin{equation}
   \label{eq:first}T_{conv} = t_{detect} + t_{prop} + t_{sync}
\end{equation}

As shown in Eq.~\ref{eq:first}, the total downtime depends upon $t_{detect}$: The time for the physical layer to report the link state change (typically $<10ms$), $t_{prop}$: The propagation delay of the Bridge Protocol Data Units across the bridge and $t_{sync}$: The time required for the handshaking process to transition a port to the Forwarding state, while the resulting data loss $P_{loss}$ (Eq.~\ref{eq:second}) is a function of the real-time traffic arrival rate during the synchronization window.

\textbf{The Data-Plane Flush:} The most significant security impact isn't only the port transition, but the MAC Address Table Flushing. Upon receiving a TCN, all switches in the spanning tree domain flush their Filtering Database (FDB) for all ports except the one receiving the TCN. The time $t_{flood}$ required to re-learn the network topology induces a unicast storm where the switch behaves like a hub:
\begin{equation}
   \label{eq:second} P_{loss} \propto \int_{t_{0}}^{t_{0}+t_{sync}} \lambda_{rt} \,dt
\end{equation}

In Eq.~\ref{eq:second}, $\lambda_{rt}$ represents the arrival rate of real-time packets. Because VoIP/Video (UDP) lacks a congestion control mechanism like TCP's $W_{cwnd}$ (congestion window), the packets are dropped without recovery, leading to the observed $2-4s$ blackout period.

The observed availability degradation results from the deterministic interaction between RSTP control-plane convergence and data-plane state invalidation. As modeled in Eq.~\ref{eq:first}, docking induced link-state changes introduce a non-zero convergence interval $T_{conv}$ during which protocol-mandated synchronization disrupts forwarding. This control-plane instability triggers topology change notifications that flush MAC address tables, causing unknown unicast flooding and packet loss. Eq.~\ref{eq:second} formalizes this loss as a function of the synchronization window and the arrival rate of latency-sensitive traffic. Because VoIP and video lack retransmission and congestion recovery, even short convergence intervals produce irreversible loss and perceptible service interruption. Although the triggering actions are legitimate and non-malicious, the resulting protocol-compliant behavior repeatedly violates availability guarantees and, consistent with NIST SP 800-61, constitutes an unintentional insider-driven denial-of-service condition.

\section{Security Implications and Mitigations}
\label{sec:mitigations}

The study demonstrates that enterprise network availability can be adversely affected by protocol compliant Layer-2 control plane behavior triggered through routine legitimate insider activity. The observed service degradation does not stem from credential misuse, misconfiguration, or policy violations, but rather from the interaction between default RSTP operation and frequent endpoint link-state changes introduced by docking and undocking peripherals. Because these events generate legitimate control-plane signaling and conform to protocol specifications, they do not manifest anomalous traffic or unauthorized actions. Consequently, the resulting availability degradation, i.e., video call freeze, remains largely undetected by conventional security monitoring systems, which are primarily designed to identify explicit attack patterns, abnormal access behavior, or traffic-based indicators of compromise.

This phenomenon also challenges current assumptions in zero trust architectures~\cite{tpm}, particularly Layer-2. Although zero-trust principles emphasize strong identity verification and least privilege access at higher layers, they often assume a stable and trusted control plane beneath them. The results of this study show that availability can be compromised even when all zero-trust assumptions are satisfied, solely through protocol-compliant behavior in the switching fabric. Consequently, availability risks originating in the control plane must be explicitly considered within enterprise threat models and risk assessments.

\subsection{Mitigation Strategy}
\label{ssec:mitigation}

To mitigate the observed availability impact, access ports connected to end hosts and docking stations were explicitly configured as edge ports using PortFast equivalent settings. This configuration ensures that such ports are treated as host only interfaces and excluded from spanning tree topology calculations. As a result, link-up and link-down events on these ports do not trigger RSTP recalculations, topology change notifications, or MAC address table flushes.

\begin{figure}[!ht]
\label{logWithoutRstp}
\centering
\includegraphics[width=1\linewidth]{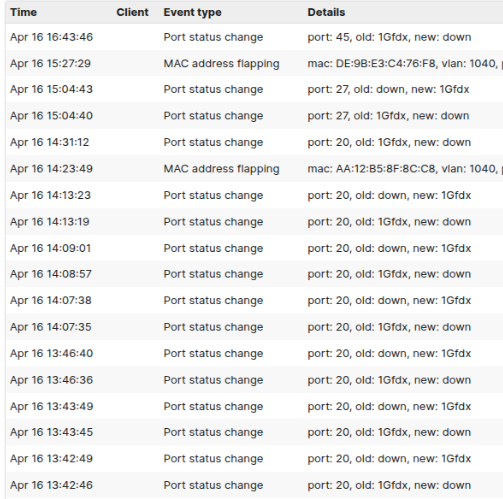}
\caption{\label{fig:norstp} Switch log status after changing the access edge port to PortFast mode.}
\end{figure}

This mitigation preserves loop-free operation while preventing benign endpoint behavior from influencing control-plane stability by explicitly excluding selected access ports from participation in the spanning tree process. Specifically, ports connected to end hosts and docking stations were reconfigured in PortFast (edge port) mode, ensuring that link-up and link-down events on these interfaces do not trigger RSTP topology changes or control plane recalculations. Only ports that were verified to connect exclusively to single host devices and that could not introduce bridging loops were transitioned to PortFast,  maintaining network safety. Following this configuration change, repeated docking and undocking events no longer induced RSTP state transitions, and no further degradation of real-time voice or video services was observed during normal operations. The proposed mitigation directly targets the terms in Eq.(~\ref{eq:first}) and Eq.(~\ref{eq:second}) by eliminating docking induced link-state changes on access ports, thereby driving the convergence interval $Tcon$ toward zero and preventing the MAC table flush that leads to real-time packet loss proportional to $ \lambda_{rt}$. Figure~\ref{fig:norstp} shows switch log status demonstrating stable port operation and absence of RSTP recalculations after configuring access interfaces in edge (PortFast) mode.

\subsection{Security Implications}
\label{ssec:securityImplications}

The effectiveness of this mitigation highlights a broader security implication, that availability protection in enterprise networks requires explicit control-plane hardening, even in environments with well-defined access control and user authentication policies. Default protocol configurations, while safe from a loop-prevention standpoint, may inadvertently expand the internal attack surface by allowing routine insider behavior to influence network-wide forwarding behavior. Without appropriate edge-port classification, this behavior can manifest itself as an unintentional, insider-driven denial-of-service condition.
These findings suggest that enterprise security practices should treat Layer-2 configuration choices as security-relevant decisions rather than purely operational optimizations. Incorporating control-plane stability considerations into security baselines and risk assessments can help prevent availability breaches that are otherwise difficult to detect, attribute, or remediate. More broadly, this work underscores the need for security frameworks to explicitly account for human-driven control-plane dynamics as a distinct class of insider threats affecting system availability.

\section{Discussion, Conclusions, and Future Work}
\label{sec:discussion}
This study analyzes availability degradation observed in a production enterprise network and is subject to limitations related to vendor-specific implementations and configuration defaults. While the behavior examined arises from standardized Layer-2 control-plane protocols, variations in hardware and software may influence the severity of impact. The analysis focuses solely on availability effects and does not consider deliberate adversarial exploitation.

Despite these limitations, the results reveal a previously underrecognized class of insider-driven availability threats. Routine, non-malicious user actions, such as frequent docking and undocking, were shown to repeatedly trigger protocol-compliant RSTP recalculations that degrade real-time enterprise services. By mapping this phenomenon to established NIST and MITRE insider threat frameworks, this work expands insider threat modeling to include unintentional actions that compromise availability without violating security policies.

These findings question assumptions about Layer-2 control-plane stability in enterprise networks and demonstrate that availability risks can arise even in well-secured environments. As a result, Layer-2 configuration choices should be treated as security-relevant decisions. Future work will focus on automating the detection of control-plane instability, validating results across additional platforms, and developing adaptive mitigation strategies aligned with modern endpoint usage patterns.

\bibliographystyle{IEEEtran}
\bibliography{sample}

\end{document}